**Photoemission kinks and phonons in cuprates.**

D. Reznik[i], G. Sangiovanni[ii], O. Gunnarsson[ii], and T. P. Devereaux[iii]

One of the possible mechanisms of high $T_c$ superconductivity is Cooper pairing with the help of bosons, which change the slope of the electronic dispersion as observed by photoemission. Giustino et al.[1] calculated that in the high temperature superconductor $La_{1.85}Sr_{0.15}CuO_4$ crystal lattice vibrations (phonons) should have a negligible effect on photoemission spectra and concluded that phonons do not play an important role. We show that the calculations employed by Giustino et al. fail to reproduce huge influence of electron-phonon coupling on important phonons observed in experiments. Thus one would expect these calculations to similarly fail in explaining the role of electron-phonon coupling for the electronic dispersion.

Density functional theory (DFT) calculations used in Ref. 1 treat electrons and phonons as independent entities, which scatter each other. Because of this scattering the electronic states acquire finite lifetimes and abrupt changes in dispersions (kinks) at the phonon energies. In addition, the phonons soften and broaden in energy. These effects are calculated from first principles without adjustable parameters. Thus if DFT is appropriate for the high $T_c$ cuprates, it ought to accurately reproduce the electronic contribution to phonon softening and broadening deduced from neutron or x-ray scattering experiments.

DFT predicts that the phonon branch in large part responsible for the calculated electronic dispersion kink is the optical bond-stretching branch involving the bond-stretching motion of $CuO_2$ plane oxygen against copper.[1] Several experimental papers highlighted large anomalous renormalization of these phonons[2-6]. They have huge low temperature dispersion dips and/or linewidth maxima around half-way (h=0.3) to the zone boundary in the superconductors $La_{1.85}Sr_{0.15}CuO_4$[2] and $YBa_2Cu_3O_7$[5] (Fig. 1). However, DFT predicts a smooth dispersion without any pronounced features in neither the dispersion nor linewidth around h=0.3 (Fig. 1). Furthermore, the very small calculated linewidths in Fig. 1b illustrate that the calculated electron-phonon coupling is very weak in absolute terms.

Substantial evidence points to electronic origin of the phonon effect. First, the phonon anomaly weakens at elevated temperatures,[2,3] whereas alternatives such as phonon-phonon scattering and structural inhomogeneity should either show the opposite trend or have no temperature dependence. Second, the phonon effect appears at specific wavevectors and is phenomenologically similar to anomalies observed in conventional systems with strong electron-phonon coupling. Third, both phonon renormalization[2] and the photoemission kink[8] become bigger when hole concentration decreases from high doping where superconductivity is suppressed towards so-called "optimal" doping with the maximum superconducting $T_c$. This simultaneous enhancement of the two features may result from an increase of electron-phonon coupling due to enhanced electronic correlations or reduced screening not included in DFT. Reference 1 cannot rule out such scenarios. The same holds for $YBa_2Cu_3O_7$ where there is a similar disagreement between the experiment and DFT results for both the phonon dispersions and the photoemission kink[9].

It is interesting that many-body calculations predict a substantial enhancement of the coupling to bond-stretching phonons compared to DFT and describe anomalous doping dependence of the zone boundary phonons[10,11], suggesting that strong correlation effects might be relevant. Recent high resolution photoemission measurements have found an oxygen isotope effect in the dispersion kink at the half-breathing phonon energy, hinting at an important role of oxygen phonons.[12] We conclude that more work is necessary to establish phonon contribution to the photoemission kink.

References:

1. Giustino, F., Cohen, M.-L. & Louie, S.-G. Small phonon contribution to the photoemission kink in the copper oxide superconductors. *Nature* **452**, 975-978 (2008).

2. Reznik, D. *et al.* Electron-phonon coupling reflecting dynamic charge inhomogeneity in copper oxide superconductors. *Nature* **440**, 1170-1173 (2006).

3. Reznik, D. *et al.* Electron-phonon anomaly related to charge stripes: Static stripe phase versus optimally doped superconducting $La_{1.85}Sr_{0.15}CuO_4$. *J. Low Temp. Phys.* **147**, 353-364 (2007).

4. Pintschovius, L. *et al.* Oxygen phonon branches in $YBa_2Cu_3O_7$. *Phys. Rev. B* **69**, 214506 (2004).

5. Uchiyama, H. *et al.* Softening of Cu-O Bond Stretching Phonons in Tetragonal $HgBa_2CuO_{4+\delta}$. *Phys. Rev. Lett.* **92**, 197005 (2004).

6. Pintschovius, L., Reznik, D. & Yamada, K. Oxygen phonon branches in overdoped $La_{1.7}Sr_{0.3}Cu_3O_4$. *Phys. Rev. B* **74**, 174514 (2006).

7. Bohnen, K.-P., Heid, R. & Krauss, M. Phonon dispersion and electron-phonon interaction for YBa2Cu3O7 from first-principles calculations. *Europhys. Lett.* **64**, 104-110 (2003).

8. Zhou, X. J. *et al.* Universal nodal Fermi velocity. *Nature* **423**, 398 (2003).

9. Heid, R. Bohnen, K.-P., Zeyher, R., & Manske, D. Momentum Dependence of the Electron-Phonon Coupling and Self-Energy Effects in Superconducting $YBa_2Cu_3O_7$ within the Local Density Approximation. *Phys. Rev. Lett*. **100**, 137001 (2008).

10. Rösch, O. & Gunnarsson, O. Electron-phonon interaction in the three-band model. *Phys. Rev. B* **70**, 224518 (2004).

11. Horsch, P. & Khaliullin, G. Doping dependence of density response and bond-stretching phonons in cuprates. *Physica B* **359-361**, 620-622 (2005).

12. Iwasawa, H. *et al.* An isotopic fingerprint of electron-phonon coupling in high-$T_c$ cuprates. Preprint at http://arxiv.org/abs/0808.1323 (2008).

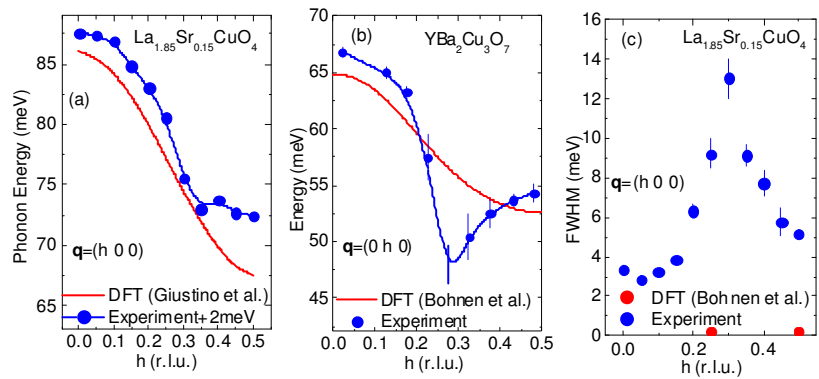

Figure 1. Comparison of some DFT predictions with experimental results for $La_{1.85}Sr_{0.15}CuO_4$ and $YBa_2Cu_3O_7$ at 10K. (a,b) Experimental bond-stretching phonon dispersions[2-4] compared to DFT results.[1,7] The data in (a) are shifted by 2meV. (c) Phonon linewidths in $La_{1.85}Sr_{0.15}CuO_4$[2,3] compared with DFT results (K.-P. Bohnen, private communication) on $YBa_2Cu_3O_7$. Ref. 1 contains no linewidth results for $La_{1.85}Sr_{0.15}CuO_4$ but we expect them to be similar. Error bars represent standard deviation. **q** represents reduced wavevector in reciprocal lattice units (r.l.u.)

[i] Forschungszentrum Karlsruhe, Institut für Festkörperphysik, P.O.B. 3640, D-76021 Karlsruhe, Germany

[ii] Max-Planck-Institut für Festkörperforschung, D-70506 Stuttgart, Germany
[iii] Dept. of Photon Science, Stanford Linear Accelerator Center, Stanford University, 2575 Sand Hill Rd., Menlo Park, CA, 94025 USA